\pdfoutput=1

\documentclass[11pt]{article}

\usepackage[preprint]{acl}

\usepackage{times}
\usepackage{latexsym}

\usepackage[T1]{fontenc}

\usepackage[utf8]{inputenc}

\usepackage{microtype}

\usepackage{inconsolata}
\usepackage{multirow}
\usepackage{booktabs}
\usepackage{graphicx}
\usepackage{amsmath}
\title{Persian MusicGen: A Large-Scale Dataset and Culturally-Aware Generative Model for Persian Music}

\author{
 \textbf{Mohammad Hossein Sameti\textsuperscript{1}},
 \textbf{Diba Hadi Esfangereh\textsuperscript{1}},
 \textbf{Sepehr Harfi Moridani\textsuperscript{1}},
 \textbf{Leili Javidpour\textsuperscript{2}},
\\
 \textbf{Mahdieh Soleymani Baghshah\textsuperscript{1}}
\\
  \textsuperscript{1}Sharif University of Technology,
  \textsuperscript{2}Independent Researcher,
\\
 \small{
   \textbf{Correspondence:} \href{mailto:soleymani@sharif.edu}{soleymani@sharif.edu}
 }
}

\usepackage{url}

\begin{document}
\maketitle
\begin{abstract}
Persian music, with its unique tonalities, modal systems (Dastgah), and rhythmic structures, presents significant challenges for music generation models trained primarily on Western music. We address this gap by curating the first large-scale dataset of Persian songs, comprising over 900 hours high-quality audio samples across diverse sub-genres, including pop, traditional, and contemporary styles. This dataset captures the rich melodic and cultural diversity of Persian music and serves as the foundation for fine-tuning MusicGen, a state-of-the-art generative music model. We adapt MusicGen to this domain and evaluate its performance by utilizing subjective and objective metrics. To assess the semantic alignment between generated music and intended style tags, we report the proportion of relevant tags accurately reflected in the generated outputs. Our results demonstrate that the fine-tuned model produces compositions that more align with Persian stylistic conventions. This work introduces a new resource for generative music research and illustrates the adaptability of music generation models to underrepresented cultural and linguistic contexts.

\end{abstract}

\section{Introduction}

Automatic music generation has emerged as a transformative frontier in artificial intelligence, democratizing musical composition and providing novel tools for creative expression, the entertainment industry, and interactive media. By lowering the technical barriers to music creation, generative systems hold the potential to augment human artistry across diverse applications. Consequently, the field has witnessed a rapid proliferation of sophisticated architectures. Recent breakthroughs, driven by diffusion and transformer-based models, have introduced powerful systems such as MusicLM~\cite{agostinelli2023musiclmgeneratingmusictext}, AudioLDM~2~\cite{liu2024audioldm}, InspireMusic \cite{zhang2025inspiremusic}, QA-MDT~\cite{li2024quality}, and FluxMusic \cite{fei2024flux}, alongside highly capable commercial platforms like Suno and Udio. These models excel at synthesizing complex, high-fidelity audio from natural language descriptions, underscoring the rapid maturation and broad utility of AI-driven music generation.

While these recent advances have led to significant improvements in automatic music generation~\cite{mehta2025musicallrepresentationalbias}, state-of-the-art models such as MusicGen employ large transformer architectures to produce high-fidelity stereo compositions from text or melody prompts~\cite{copet2024simplecontrollablemusicgeneration}. However, these systems typically rely on vast Western-centric music corpora and may fail to capture the musical idioms of underrepresented cultures. In practice, generative models often reflect these biases: for example, only about 5.7\% of the hours in existing music datasets come from non-Western traditions~\cite{mehta2025musicallrepresentationalbias}, and models trained on such data tend to impose Western tonal and rhythmic patterns when generating music in other styles. This disparity suggests that music generation requires culturally-aware methods capable of learning from limited data in low-resource settings~\cite{mehta2025exploringadapterdesigntradeoffs}.

Persian music exemplifies such a low-resource, culturally distinct case. It is built on the classical \textit{Dastgah} modal system: each Dastgah comprises a sequence of \textit{gusheh} (melodic motifs) organized around a modal center~\cite{babak_nikzat_2022_7316660}. Unlike Western scales, Persian modes include flexible microtonal intervals (smaller than a semitone) that vary by gusheh. Performances blend free-form ornamentation with structured rhythms: vocal sections often follow Persian poetic meter (\textit{aruz}) in free tempo, while instrumental sections may employ cyclical rhythms (e.g., \textit{kereshmeh}, \textit{chahārpareh})~\cite{babak_nikzat_2022_7316660}. These unique melodic and rhythmic features highlight why naïvely applying Western-trained models to Persian music can produce musically incongruous results.

Compounding these challenges, there is a severe lack of digital resources for Persian music. For instance, \cite{jafari2024vocalmelodyconstructionpersian} had to collect and transcribe over a hundred Persian songs from print because no suitable dataset existed. The few publicly available corpora focus almost exclusively on classical and traditional forms. For example, the Nava dataset contains 1,786 traditional performances by 40 artists~\cite{Nava}, and the KDC corpus has 213 complete Dastgāh performances~\cite{babak_nikzat_2022_7316660}. Other recent efforts, while valuable, continue this focus on classical music, such as the IRMA dataset of audio-MIDI radif performances~\cite{Shafiei_2025} and the symbolic Radif Corpus~\cite{kanani2025radifcorpussymbolicdataset}. Crucially, this leaves modern Persian pop, folk, and fusion genres, which represent a vast portion of the culture's music, entirely unrepresented. In short, the absence of a comprehensive, large-scale, and stylistically diverse Persian music dataset has hindered generative research in this area.

To address these gaps, we make three key contributions: 
\begin{enumerate}
    \item \textbf{Dataset curation:} We compile a large-scale Persian music dataset spanning pop, folk, and contemporary genres, filling a critical resource gap.
    \item \textbf{Model adaptation:} We fine-tune the MusicGen model on this dataset, adapting a music generation system to Persian musical traditions.
    \item \textbf{Comprehensive evaluation:} We analyze the generated samples to assess \textit{cultural fidelity} and overall musical quality, providing both quantitative and human-centered evaluation of how well the model captures Persian musical characteristics.
\end{enumerate}

These contributions lay the groundwork for culturally-aware generative music models in low-resource settings, demonstrating how modern generative architectures can be adapted to the rich modalities of Persian music.
\section{Related Work}


\subsection{Music Generation Datasets}

\begin{figure*}[t]
  \centering
  \includegraphics[width=\linewidth]{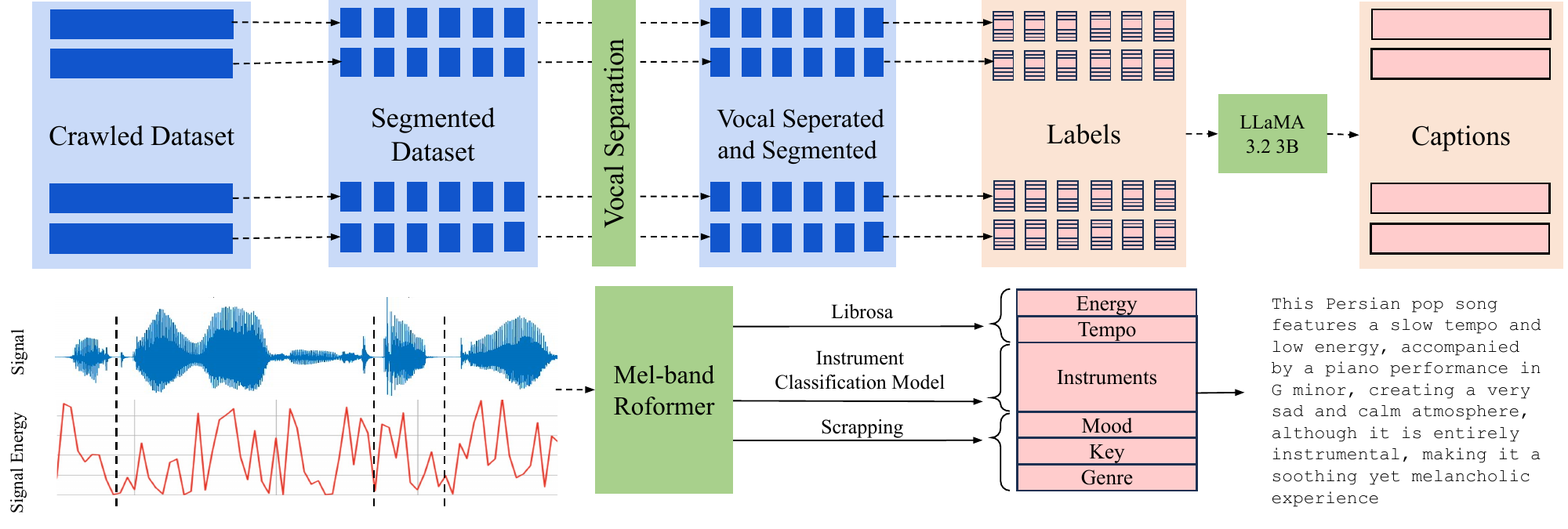}
  \caption{Overview of the dataset creation pipeline, consisting of data crawling, segmentation, source separation, tagging, and caption generation.}
  \label{fig:dataset-pipeline}
\end{figure*}

Recent advances in music generation have been driven by the availability of large-scale symbolic and audio datasets, primarily focused on Western music traditions. For instance, MAESTRO is a widely used dataset of aligned audio and MIDI recordings of classical piano performances, enabling supervised training on precise note timings and dynamics\cite{maestro}. The Lakh MIDI Dataset, one of the largest symbolic music corpora, comprises over 170,000 MIDI files derived from Western popular music, spanning genres like rock, jazz, and pop\cite{lakhmidi}. Slakh2100 expands on this by synthesizing multitrack audio renditions from Lakh MIDI files using high-quality instrument samples, thereby facilitating research in audio-domain music generation with multi-instrumental structure\cite{slakhmidi}.

Other datasets such as MusicNet and the Free Music Archive (FMA) provide aligned audio and metadata for classical and Creative Commons-licensed music, respectively, further broadening the training resources for deep generative models\cite{musicnet, fma_dataset}. Similarly, the MTG-Jamendo dataset provides over 55,000 full-length, high-quality tracks with rich genre, instrument, and mood tags, becoming a key public benchmark for audio-domain tasks~\cite{mtg_jamendo}. Recently, XMIDI has emerged as a large-scale dataset with fine-grained emotion and genre annotations across over 100,000 MIDI files, enhancing control and conditioning in music generation tasks\cite{xmidi}. Despite their utility, these datasets share a common limitation: they overwhelmingly represent Western musical structures, genres, and instrumentation. Consequently, models trained on them tend to replicate Western tonalities and rhythmic patterns, which limits their applicability to underrepresented musical traditions.
\subsection{Bias in Music Datasets and Models}
A critical limitation of the mentioned datasets in Section 2.1 is their overwhelming Western bias. This issue was quantified in a study, which analyzed over one million hours of music corpora. The study found that nearly 94\% of the data represents Western music, with only about 5.7\% coming from all other non-Western genres combined. Crucially, traditions like Hindustani, Middle Eastern, and African music accounted for a tiny fraction of the data (e.g., Middle Eastern traditions at under 1\%). This profound data imbalance has a direct, negative impact on model performance: state-of-the-art generative models trained on such data tend to default to Western tonal and rhythmic conventions, even when explicitly prompted for other styles~\cite{mehta2025musicallrepresentationalbias}.
\subsection{Culturally-Specific and Persian Music Data}
Only a few datasets focus on non-Western or low-resource musical traditions. For Persian (Iranian) music specifically, the most notable resources focus on classical and traditional forms. The Nava dataset, for instance, contains 1,786 solo recordings (55 hours) of five traditional instruments (kamancheh, tar, setar, etc.) played by 40 artists across the seven main Dastgah systems~\cite{ebrat2022iranianmodalmusicdastgah}. Another collection is the Persian Piano Corpus (PPC), which includes 2,022 piano pieces from Persian composers, annotated with their Dastgah modes~\cite{rasouli2023persianpianocorpuscollection}. These datasets have enabled research on tasks like Dastgah classification and instrument recognition but are modest in size and highly genre-specific.

This focus on classical forms is consistent across other recent efforts. For example, the IRMA dataset provides an audio-MIDI corpus of radif performances~\cite{Shafiei_2025}, and the symbolic Radif Corpus offers a dataset for non-metric Iranian classical music~\cite{kanani2025radifcorpussymbolicdataset}. Smaller, specialized corpora also follow this pattern, such as a dataset of ~700 short clips of seven Persian classical instruments collected for an instrument recognition study~\cite{mousavi2019persian}. Beyond these, only very small or unpublished corpora of Persian music exist (e.g., a 500-song genre-labeled set). Importantly, no large-scale, public dataset of contemporary Persian music (modern pop, rock, rap, etc.) is available to our knowledge. This stands in stark contrast to the plentiful Western data, and it highlights a major gap: modern Persian musics remain essentially absent from current generative models’ training data.

\begin{table*}[t]
\centering
\footnotesize
\caption{Summary statistics of the Persian music dataset (67,796 tracks).}
\label{tab:dataset_stats}

\begin{tabular}{@{}lll@{}}
\toprule
\multicolumn{3}{@{}l@{}}{\textbf{Musical Characteristics}} \\
\midrule

\begin{tabular}[t]{@{}lr@{}}
\textbf{Key Distribution} & \\
\midrule
\textbf{Key} & \textbf{Count} \\
\midrule
B Minor & 6905 \\
E Minor & 6248 \\
A Minor & 5920 \\
C Minor & 5680 \\
D Minor & 5394 \\
C$\sharp$/D$\flat$ Min & 5298 \\
A$\sharp$/B$\flat$ Min & 4519 \\
F Minor & 4509 \\
G Minor & 3932 \\
F$\sharp$/G$\flat$ Min & 3773 \\
D$\sharp$/E$\flat$ Min & 3460 \\
G$\sharp$/A$\flat$ Min & 2803 \\
C Major & 1924 \\
G Major & 1653 \\
C$\sharp$/D$\flat$ Major & 1213 \\
F$\sharp$/G$\flat$ Major & 1139 \\
F Major & 677 \\
A Major & 621 \\
G$\sharp$/A$\flat$ Major & 617 \\
D Major & 502 \\
A$\sharp$/B$\flat$ Major & 372 \\
B Major & 289 \\
D$\sharp$/E$\flat$ Major & 176 \\
E Major & 172 \\
\end{tabular}
&
\begin{tabular}[t]{@{}lr@{}}
\textbf{Tempo Distribution} & \\
\midrule
\textbf{Tempo} & \textbf{Count} \\
\midrule
Moderate & 43317 \\
Slow & 21430 \\
Upbeat & 2641 \\
Fast & 408 \\
\\
\\
\textbf{Energy Distribution} & \\
\midrule
\textbf{Energy} & \textbf{Count} \\
\midrule
Moderate & 56353 \\
Low & 9905 \\
High & 1538 \\
\\
\\
\textbf{Top Instruments} & \\
\midrule
\textbf{Instrument} & \textbf{Count} \\
\midrule
Synthesizer & 6342 \\
Piano & 5310 \\
Ney & 2320 \\
Acoustic Guitar & 2033 \\
Bass & 853 \\
Drums & 706 \\
Daaf & 623 \\
Kamancheh & 599 \\
Tonbak & 517 \\
Setar & 427 \\
Santur & 276 \\
Tar & 250 \\
\end{tabular}
&
\begin{tabular}[t]{@{}lr@{}}
\textbf{Genre Distribution} & \\
\midrule
\textbf{Genre} & \textbf{Count} \\
\midrule
Persian pop & 63531 \\
Persian rock & 3276 \\
Classic persian pop & 2299 \\
Persian traditional & 1277 \\
Persian alternative & 232 \\
Neo-traditional & 11 \\
\\
\textbf{Happiness Distribution} & \\
\midrule
\textbf{Range} & \textbf{Count} \\
\midrule
0--9 & 603 \\
10--19 & 3016 \\
20--29 & 5965 \\
30--39 & 9793 \\
40--49 & 10235 \\
50--59 & 10354 \\
60--69 & 9552 \\
70--79 & 8269 \\
80--89 & 6546 \\
90--99 & 3463 \\
\\
\textbf{Popularity Distribution} & \\
\midrule
\textbf{Popularity} & \textbf{Count} \\
\midrule
0--9 & 31306 \\
10--19 & 16899 \\
20--29 & 11758 \\
30--39 & 6568 \\
40--49 & 1241 \\
50--59 & 24 \\
\end{tabular}
\\
\bottomrule
\end{tabular}
\end{table*}

\section{Dataset}
As described in the previous sections, the first step is to construct a rich dataset of Iranian music by developing a comprehensive pipeline for extracting audio pieces along with their corresponding textual annotations. Figure~\ref{fig:dataset-pipeline} illustrates the overall pipeline of dataset creation and the details of each component of this pipeline will be explained in the following sections. Additionally, the distribution of the extracted features across all collected samples is summarized in Table \ref{tab:dataset_stats}, providing an overview of the dataset’s statistical characteristics.\footnote{\url{https://huggingface.co/datasets/mohammadhossein/PMG}}

\begin{figure*}[t]
  \centering
  \includegraphics[width=\linewidth]{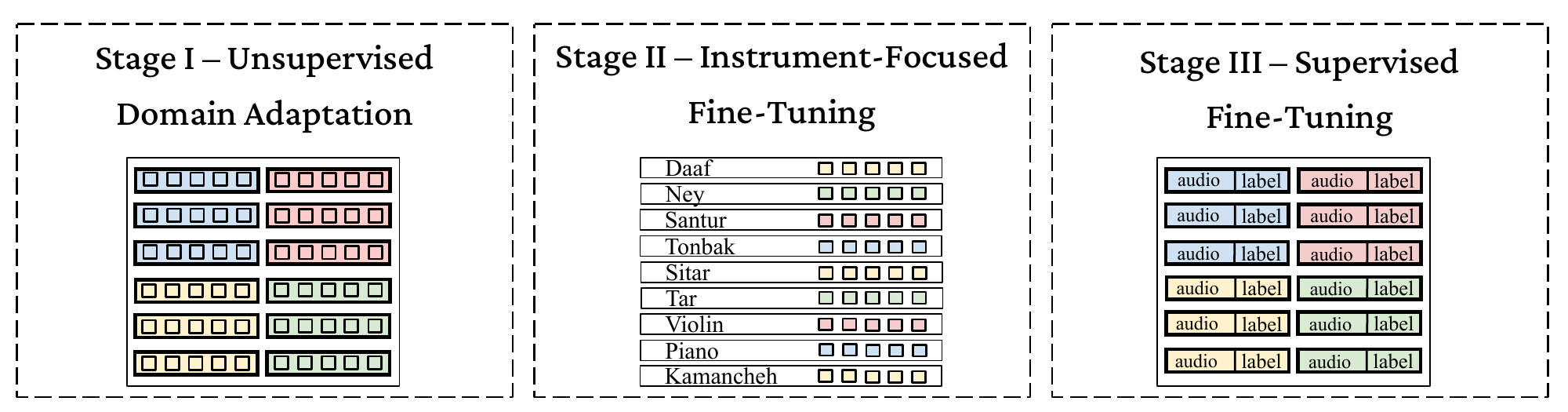}
  \caption{Overview of the training pipeline.}
  \label{fig:training-pipeline}
\end{figure*}

\subsection{Crawling Persian Music Tracks}
To build a large-scale dataset of Persian music, we crawled audio tracks from a prominent Persian music website that hosts a comprehensive collection of contemporary and classical works. The platform contains a rich archive of songs across various genres, moods, and artists. In total, we collected approximately 16k tracks of Persian music, covering a wide temporal and stylistic spectrum. This large corpus served as the foundation for further processing, segmentation, and annotation steps in our dataset pipeline.
\subsection{Segmenting Tracks into Musical Clips}
To prepare the music data for model training, we segmented each Persian track into shorter clips ranging from 10 to 30 seconds in duration. Instead of fixed-length slicing, we used an adaptive segmentation strategy based on signal energy. Signal energy was analyzed over short overlapping frames to identify points of significant dynamic change, such as the onset or offset of a musical phrase.  Segmentation boundaries were chosen where this features indicated structural changes, helping preserve musical coherence within each clip. This method allowed us to capture meaningful musical units aligned with the natural phrasing of Persian music while maintaining consistency in clip length.
\subsection{Tag Prediction}
Updated Technical Description
To enrich the metadata associated with each musical clip and enable more expressive conditional generation, we performed automatic tag prediction using a modular approach. For tempo and energy, we relied on signal analysis features computed with the Librosa library. Tempo was estimated using Librosa’s beat tracking utilities, which compute the beats-per-minute (BPM) based on rhythmic periodicity in the signal. Energy was calculated by analyzing the short-time signal amplitude across frames, and clips were then categorized into coarse energy levels (e.g., low, medium, high), capturing the overall dynamic intensity.

For instrument classification, we employed a bifurcated strategy based on the musical genre:
\begin{itemize}
    \item Traditional Persian Tracks: We utilized the specialized model introduced by \cite{esfangereh2025persian}, which is optimized for the nuanced timbral characteristics of Iranian instruments.
    \item General/Other Tracks: For non-traditional clips, we implemented the model described in \cite{yuan2023marble}, leveraging its robust performance in broad-spectrum instrument tagging.
\end{itemize}

This step was crucial for Persian music, where traditional instruments play a central role in musical expression. The combination of these tags formed a structured set of semantic labels used to condition the generation process. This multi-aspect tagging strategy enabled more informed and musically coherent generation aligned with the unique characteristics of Persian music.
\subsection{Vocal-Instrumental Separation}
To enable finer control over the musical attributes and improve conditioning for generation, we performed vocal-instrumental separation on all clips. We utilized the Mel-band Roformer model, a recent state-of-the-art method for music source separation\cite{wang2023melbandroformermusicsource}. Unlike traditional band-split approaches that divide the frequency spectrum into non-overlapping subbands heuristically, Mel-band Roformer uses a perceptually grounded mel-scale projection. The model converts the audio into a complex-valued mel-spectrogram, where frequency bins are mapped into overlapping mel bands, reflecting the human ear’s nonlinear perception of pitch. A hierarchical Transformer with Rotary Position Embeddings (RoPE) is then used to model both intra- and inter-band dependencies across time, capturing the long-range structure crucial for music. The model estimates a mask for each mel band, which is then projected back to the full frequency domain for waveform reconstruction. This approach has shown superior performance in separating vocals, drums, and other stems compared to prior models like BSRNN or BS-RoFormer, especially in terms of signal-to-distortion ratio (SDR)\cite{bsrnn, bsroformer}. In our pipeline, we leveraged this capability to extract clean vocal and instrumental stems from Persian music clips, which were later used for tag prediction and conditional generation.
\subsection{Caption Generation using LLaMA 3.2 3B}

To create natural language descriptions for each musical clip, we used the LLaMA 3.2 3B large language model to generate rich, human-like captions using the dataset's tags\cite{llama3}. These captions serve as semantic summaries of the musical content, aiding both dataset exploration and potential text-to-music generation tasks. Each prompt to the model included structured metadata derived from previous steps, namely, the predicted instruments, tempo, and energy level, along with additional mood, key, and genre labels scraped from online sources (e.g., song databases or streaming metadata). The input to LLaMA was formatted as a concise prompt template, and the model was prompted in zero-shot mode to produce fluent and contextually relevant captions.

To further enhance the diversity and expressiveness of the generated descriptions, we applied several augmentation strategies during prompt construction: random rotation of tag order, random omission of selected tags, and the incorporation of additional general information about the singer, their stylistic tendencies, and the typical atmosphere of their songs. These variations reduced template rigidity and encouraged the model to generate more varied, natural, and stylistically appropriate captions. Overall, this procedure yielded expressive textual descriptions that closely aligned with the musical attributes of Persian music, significantly enriching the semantic quality and usability of the dataset.

\section{Training Procedure}

We base our approach on \textit{MusicGen}, a powerful and publicly available generative model designed for text-conditioned music synthesis\cite{copet2024simplecontrollablemusicgeneration}. MusicGen employs a single-stage Transformer decoder that autoregressively generates discrete audio tokens interleaved with optional text-conditioning tokens. It leverages a high-quality audio tokenizer (e.g., EnCodec) to represent stereo audio as discrete token streams, allowing the model to operate in a language-modeling framework. The resulting system is capable of generating coherent musical audio from either free-form text prompts or melody conditioning.

To effectively adapt MusicGen to the Persian music domain, it is essential not only to control the discrete audio tokens in a manner that faithfully reflects the characteristics of Persian musical pieces, but also to ensure coherent alignment with their textual descriptions. To this end, we introduce a three-stage training pipeline that guides the model from general audio generation toward culturally specific, instrument-aware Persian composition conditioned on fine-grained textual details. An overview of this training procedure is presented in figure \ref{fig:training-pipeline}.

\subsection{Stage I – Unsupervised Domain Adaptation}

In the first phase, we utilize our large-scale Persian music dataset, including over 900 hours audios across traditional, pop, and contemporary genres, to perform \textit{self-supervised domain adaptation}. 
This stage fine-tunes only the autoregressive language model component of MusicGen on unlabeled Persian audio, enabling it to internalize modal, rhythmic, and melodic regularities characteristic of the Dastgāh system. 
Through exposure to domain-specific token distributions, the model acquires an implicit representation of Persian tonal organization and microtonal intervals. 
This unsupervised alignment corresponds to the principle of domain-specific pre-adaptation, where a pretrained generative prior is gradually specialized toward a target distribution without requiring paired supervision.

\subsection{Stage II – Instrument-Focused Fine-Tuning}

In the second phase, we refine the model on a curated subset of \textbf{solo-instrument recordings} (e.g., tar, setār, santūr, kamancheh, daf) \cite{esfangereh2025persian}. 
These monophonic and low-polyphonic excerpts allow the model to disentangle timbral and modal cues from accompaniment noise. 
Training on instrument-isolated data promotes \textbf{factorized representation learning} of timbre and pitch contour, a property theoretically supported by research on disentangled audio representations. 
This intermediate step improves the model’s ability to reproduce authentic Persian instrument tones and ornamentations, essential features of expressive style that are often blurred in full-mix training.

\subsection{Stage III – Supervised Fine-Tuning with Paired and Solo Data}

Finally, we perform \textit{supervised fine-tuning} using a smaller paired subset containing text-audio examples such as descriptive captions (e.g., “melancholic Persian pop with santur intro”) and stylistic tags. 
We jointly optimize both the language model and text-conditioning modules using the combined multi-instrument and solo-instrument data. 
This phase reinforces two complementary capabilities:

\begin{itemize}
    \item \textbf{Semantic Alignment:} The text-conditioning encoder becomes aligned with Persian musical semantics, including Dastgāh modes, timbral elements, and culturally salient rhythmic motifs, improving coherence between prompts and outputs.
    \item \textbf{Musical Structure:} Continued decoder training with aligned and timbre-refined data yields more structured, stylistically faithful compositions.
\end{itemize}


\section{Experiments \& Results}

\begin{table*}[h]
\centering
\caption{Performance comparison of Our Model and the MusicGen Baseline across Traditional (Polyphonic, Solo) and Pop Persian music generation.}
\label{tab:perf_all}
\renewcommand{\arraystretch}{1.35}

\begin{tabular}{lcccccc}
\toprule
\multirow{2}{*}{\textbf{Model}} 
& \multicolumn{4}{c}{\textbf{Traditional}} 
& \multicolumn{2}{c}{\textbf{Pop}} \\
\cmidrule(lr){2-5} \cmidrule(lr){6-7}
& \multicolumn{2}{c}{\textbf{Monophonic}} 
& \multicolumn{2}{c}{\textbf{Polyphonic}} 
& \multirow{2}{*}{\textbf{KLD}} & \multirow{2}{*}{\textbf{Chroma}} \\
\cmidrule(lr){2-3} \cmidrule(lr){4-5}
& \textbf{KLD} & \textbf{Chroma} 
& \textbf{KLD} & \textbf{Chroma}
&  &  \\
\midrule

\textbf{Our Model} 
& 5.28 & 0.40 
& 3.23 & 0.44
& 3.64 & 0.51 \\

\textbf{MusicGen (Baseline)} 
& 6.37 & 0.33
& 3.43 & 0.36 
& 4.27 & 0.46 \\

\bottomrule
\end{tabular}
\end{table*}

\begin{table*}[h]
\centering
\caption{Comparison of chroma similarity across three evaluation subsets under four conditioning schemes: text-only (\(\mathcal{C}_{\text{text}}\)) and text combined with 1s, 3s, and 5s of audio context.}
\label{tab:chroma_similarity_results}

\begin{tabular}{|c|c|c|c|c|c|c|}
\hline
\multirow{2}{*}{\textbf{Condition}} & 
\multicolumn{2}{c|}{\textbf{Multi Instruments}} &
\multicolumn{2}{c|}{\textbf{Solo Instruments}} &
\multicolumn{2}{c|}{\textbf{Pop}} \\
\cline{2-7}
 & \textbf{Ours} & \textbf{Base} & \textbf{Ours} & \textbf{Base} & \textbf{Ours} & \textbf{Base} \\
\hline
\(\mathcal{C}_{\text{text}}\) & 0.4471 & 0.3689 & 0.4011 & 0.3316 & 0.5115 & 0.4663 \\
\hline
\(\mathcal{C}_{\text{text}+1\mathrm{s}}\) & 0.5059 & 0.4813 & 0.4565 & 0.4353 & 0.5781 & 0.5625 \\
\hline
\(\mathcal{C}_{\text{text}+3\mathrm{s}}\) & 0.5644 & 0.5470 & 0.5050 & 0.5148 & 0.6208 & 0.6359 \\
\hline
\(\mathcal{C}_{\text{text}+5\mathrm{s}}\) & 0.6131 & 0.6037 & 0.5816 & 0.5896 & 0.6735 & 0.6855 \\
\hline
\end{tabular}

\end{table*}

Our experimental evaluation investigates the effectiveness of the adapted MusicGen model in generating culturally coherent Persian music. The small variant of MusicGen was fine-tuned on our curated dataset spanning multiple genres, including traditional, pop, and contemporary styles, to ensure broad stylistic coverage.

We evaluate model performance using two objective metrics:
(1) Kullback–Leibler Divergence (KLD), which measures the divergence between feature distributions in generated and authentic Persian music, and
(2) Chroma Cosine Similarity, which quantifies harmonic consistency between generated audio and reference pieces.
As reported in Table~\ref{tab:perf_all}, our model achieves lower KLD and higher chroma similarity than the baseline, indicating improved alignment with Persian musical structure and harmonic patterns.


We additionally examine the effect of different conditioning strategies on the harmonic fidelity of the generated music. Four configurations are considered (see Table~\ref{tab:chroma_similarity_results}): \(\mathcal{C}_{\text{text}}\), and three text-plus-audio variants that incorporate 1, 3, or 5 seconds of prefix audio \(\mathcal{C}_{\text{text}+1\mathrm{s}}\), 
\(\mathcal{C}_{\text{text}+3\mathrm{s}}\),
\(\mathcal{C}_{\text{text}+5\mathrm{s}}\). Increasing the duration of audio conditioning provides the model with richer cues regarding timbre, tuning, and harmonic structure. Consistent with this intuition, harmonic similarity improves progressively with longer audio prefixes. This trend is observed across all evaluation subsets, specifically multi-instrument recordings, solo-instrument tracks, and Persian pop pieces, which highlights the utility of hybrid text–audio conditioning for Persian music generation.




\section{Conclusion}
This study presents the curation of the first large-scale, high-quality Persian music dataset and demonstrates the successful adaptation of the MusicGen model to this culturally rich domain. By fine-tuning MusicGen on this comprehensive corpus which comprises over 900 hours of audio across pop, traditional, and contemporary genres, we have shown that it is possible to adapt state-of-the-art generative models to culturally distinct and historically underrepresented musical traditions.

Our experimental results confirm that the fine-tuned MusicGen model produces music significantly more aligned with Persian musical structures. This is evidenced by lower KLD values and higher Chroma Cosine Similarity compared to the baseline, indicating that the model successfully captured the unique modal and rhythmic idioms of the Dastgah system. These results provide a promising avenue for future research into culturally-aware music generation.

In future work, we plan to further enhance the model's capabilities by incorporating additional musical features and extending the dataset to include even more genres. We also aim to explore more advanced fine-tuning techniques to improve the model's ability to generate music with even greater stylistic accuracy

\section{Limitations}
\paragraph{Dataset Scope and Balance.} Although our dataset comprises over 900 hours of Persian music, its genre distribution is heavily skewed toward Persian pop, which accounts for approximately 93.7\% of all tracks. Genres such as traditional, rock, and neo-traditional music are substantially underrepresented. This imbalance may bias the model toward pop-style generation and limit its ability to faithfully reproduce the melodic and rhythmic nuances of classical and traditional Persian music, where the Dastgah system is most prominently expressed.
\paragraph{Automatic Tagging and Caption Quality.} The metadata used for conditioning were extracted using automated pipelines rather than expert human annotation. These automatic predictions inevitably introduce noise; for example, instrument classifiers may confuse timbral similarities between Persian instruments (e.g., tar and setar), and key estimation tools designed for Western tonalities may not reliably capture Persian microtonal intervals. Similarly, the captions generated by LLaMA 3.2 3B, while fluent, were not systematically validated by domain experts, potentially introducing semantic inaccuracies in the text conditioning.
\paragraph{Evaluation Methodology.} Our evaluation relies primarily on KL Divergence and Chroma Cosine Similarity, both of which measure distributional and harmonic alignment but do not directly assess perceptually salient qualities of Persian music such as microtonal accuracy, ornamentation fidelity, or adherence to Dastgah modal progressions. Furthermore, while we employ a hybrid human–classifier evaluation for tag correctness, a comprehensive subjective listening study with trained Persian musicians, assessing cultural authenticity, emotional expressiveness, and stylistic coherence, was not conducted. Such expert evaluation would provide a more meaningful assessment of the model's cultural fidelity.
\paragraph{Model Scale.} We fine-tuned only the small variant of MusicGen due to computational constraints. Larger model variants may offer improved capacity for capturing the complex melodic structures and long-range dependencies characteristic of Persian music, particularly in traditional forms where extended improvisatory passages (avaz) are common. The generalizability of our findings to larger architectures remains unexplored.
\paragraph{Microtonal and Modal Representation.} The underlying audio tokenizer (EnCodec) and the MusicGen architecture were originally designed for Western music, which operates on a 12-tone equal temperament system. Persian music relies on microtonal intervals that fall between Western semitones and vary contextually across different Dastgah modes. It is unclear to what extent the current tokenization scheme can faithfully encode and reconstruct these subtle pitch distinctions, and no dedicated analysis of microtonal fidelity was performed.

\bibliography{main}

\end{document}